\documentclass[]{article}

\usepackage{booktabs}
\usepackage{graphicx}
\usepackage{url}

\begin{document}

\title{(In)Secure Configuration Practices\\of WPA2 Enterprise Supplicants}

\author{
    Alberto Bartoli, Eric Medvet, Andrea De Lorenzo, Fabiano Tarlao\\
    University of Trieste, Italy\\
    \texttt{\{bartoli.alberto, emedvet, andrea.delorenzo, ftarlao\}@units.it}
}

\maketitle

\textbf{Please cite as:} Alberto Bartoli, Eric Medvet, Andrea De Lorenzo, and Fabiano Tarlao. 2018. (In)Secure Configuration Practices of WPA2 Enterprise Supplicants. In \textit{Proceedings of Availability, Reliability and Security, Hamburg, August 2018 (ARES), 6 pages}

\begin{abstract}
WPA2 Enterprise is a fundamental technology for secure communication in enterprise wireless networks. 
A key requirement of this technology is that WiFi-enabled devices (i.e., \emph{supplicants}) be correctly configured before connecting to the enterprise wireless network.
Supplicants that are not configured correctly may fall prey of attacks aimed at stealing the network credentials very easily.
Such credentials have an enormous value because they usually unlock access to \emph{all} enterprise services.

In this work we investigate whether users and technicians are aware of these important and widespread risks.
We conducted two extensive analyses: a survey among approximately 1000 users about how they configured their WiFi devices for enterprise network access; and, a review of approximately 310 network configuration guides made available by enterprise network administrators.
The results provide strong indications that the key requirement of WPA2 Enterprise is violated systematically and thus can no longer be considered realistic. 
\end{abstract}


\section{Introduction}

Secure WiFi access is a fundamental component of every enterprise.
This technology is based on the WPA2 Enterprise family of protocols and ensures that only authenticated devices may connect to an enterprise wireless network: each user has personalized credentials and devices may connect only after presenting valid credentials of a user \cite{Les_Owens_Karen_Scarfone2007-lt}.
After successful authentication, the wireless traffic between the device (\emph{supplicant} in WPA2 Enterprise parlance) and the access point occurs with strong security guarantees: secrecy, integrity and mutual authentication.
The protocols for providing these guarantees are such that devices owned by different users need not trust each other.
As such, WPA2 Enterprise provides a strong and secure foundation for wireless communication within any large organization.

A key requirement of WPA2 Enterprise is that supplicants be correctly configured before connecting to the enterprise wireless network \cite{Souppaya2012-nk}.
While the exact meaning of correctly configured may depend on the specific environment, a fundamental requirement is that the supplicant knows in advance the DNS name of the \emph{authorization server} of the enterprise network, that is, of the server that validates the credentials exhibited by a supplicant during the authentication that must occur prior to connecting.

Satisfying this requirement is extremely important.
Supplicants that are not correctly configured, i.e., that do not know the DNS name of the authorization server for the enterprise network, may be easily tricked into starting an execution of the authentication protocol with an \emph{evil twin}, a malicious access point that broadcasts the SSID of the enterprise network (e.g., \cite{Yavor_undated-kk,noauthor_2013-gh,Cassola2013-ox,Snoodgrass_J_undated-kf,Brenza2015-mm}).
In most configurations the protocol execution will fail, because the evil twin will not be able to authenticate itself to the supplicant as required by WPA2 Enterprise.
However, even a failed execution may suffice to leak user credentials to the evil twin---the supplicant may decide to abort execution when it is too late, i.e., after it has already sent credential material to the evil twin.

This fact is not a merely hypothetical risk.
It has been recently showed in \cite{Bartoli2018-pd} that by just wandering around for a few hours in regions not covered by a wireless network one could collect 200 enterprise credentials---roughly 25\% of them in clear text and the remaining ones in hashed MS-CHAPv2 form, which can generally be decrypted easily \cite{noauthor_2012-dj}.
It was also showed that by remaining for a few seconds at less than 35 meters from a specific (voluntary) target whose wifi device is not configured correctly, there is a very good chance that you may steal his/her enterprise credentials; even when he/she is sitting in a car with close windows.
We are thus facing a scenario with attacks that can be done quickly, cheaply, everywhere outside of the enterprise, that do not require any explicit action from the targeted user; attacks that deliver, if successful, the enterprise credentials, either in the clear or in a format that may be attacked offline with moderate effort.
We emphasize that we are not concerned with man-in-the-middle attacks (MITM), i.e., attacks in which the supplicant connects to the Internet through a fraudulent access point.
We are concerned with attacks consisting in stealing network credentials by means of an execution of the network authentication protocol: these attacks may occur in an few seconds and without any involvement of the user, unlike MITM attacks.

What makes this issue particularly relevant and important is the combination of two factors: nowadays most enterprises are based on single sign on, i.e., the same credentials that are used for authenticating to the wireless network unlock access to \emph{all} the services of the enterprise; and, virtually everyone is equipped with a portable WiFi device (a smartphone) that is permanently attempting to connect to known SSIDs.

In this work we aim at gaining some insights into whether users and technicians are really aware of these important and widespread risks. Anecdotal evidence and our own experience suggest that this is not the case.
For this reason, we conducted two extensive analyses: a survey among approximately 1000 users from
many different institutions about how they configured their WiFi devices for enterprise network access;
and, a review of approximately 310 network configuration guides made available by enterprise network administrators.

The results of our analyses, illustrated in the next sections, provide important insights into the practical deployments of such a pervasive and fundamental technology as WPA2.
Perhaps most importantly, these results show that the basic security assumption of WPA2 Enterprise \emph{``you are expected to configure your device appropriately, otherwise you will still connect to an authenticated access point but you might be connected with an attacker''} is indeed violated systematically and thus can no longer be considered realistic. 
This is an extremely important point to make, because the WiFi Alliance is defining a new generation of secure protocols for wireless networking (WPA3) that will address several shortcomings of the current WPA2 standard (e.g., the recently discovered KRACK vulnerability \cite{Vanhoef2017-gb}).
The publicly available information on WPA3, though, does not address the issue of supplicant configuration at all \cite{noauthor_2018-ty}.
We are thus incurring the risk of receiving a novel fundamental standard constructed upon requirements that in many cases are not satisfied, thereby leading to practical deployments of the standard that in many cases will not be secure.

\section{Survey of configuration practices}

In order to gain insights into how users actually configure their wifi devices,
we circulated a survey in the \emph{eduroam} user community.
Eduroam is a roaming access service provided by a number of research
institutions \cite{Wierenga2015-vm}.
It allows members of participating institutions to obtain Internet connectivity
at any other participating institution.
In 2016, eduroam provided over 2.6 billion authentications of roaming users in the same country and over 592 million cross-border authentications\footnote{https://www.eduroam.org/2017/03/07/2016-a-record-breaking-year-for-eduroam/}. 

Focussing on eduroam is important for several reasons.
\begin{itemize}
\item Each user should connect to eduroam according to WPA2 Enterprise configuration instructions specific
to his/her home institution. Thus, the survey allowed collecting data about the practices followed at many different organizations.
\item Eduroam has developed an application available for all the major operating systems that allows users to configure their devices automatically, with the configuration suitable for their home institution---\emph{eduroam-cat}\footnote{https://cat.eduroam.org/}.
Thus, the survey allowed collecting data about the actual behavior of users, i.e., whether they are able or willing to configure their devices autonomously, or they resort to the tool for automatic configuration.
We emphasize that usage of the tool ensures a secure configuration of the supplicant while autonomous configuration does not.
\item Eduroam is a federation of research institutions, thus their users may be more willing to share details about their technical practices than corporate environments.
\end{itemize}
Indeed, we are not aware of any similar survey regarding the practices actually followed by users for configuring their WPA2 Enterprise supplicants (see also below).

The survey consisted of a few non-technical questions to be answered on a Google Form, available in Italian and in English.
We used our personal contacts as initial seed.
We collected 1099 answers, of which 964 from users which had connected at least one of their wifi devices to eduroam.
In the following we will consider only answers from these users.
We submitted several questions separately for smartphone, tablet, notebook and obtained answers for 2054 devices.

\begin{figure}
    \centering
    \includegraphics[width=1\linewidth]{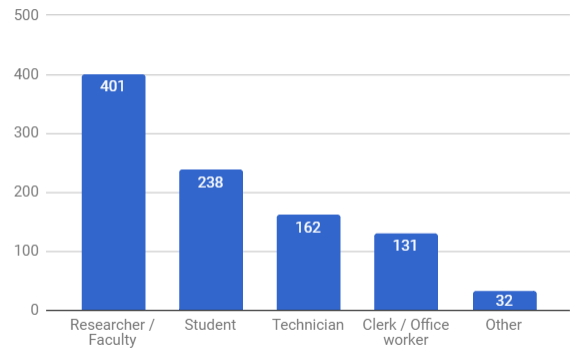}
    \caption{
        \label{fig:user-breakdown}
        Role of users participating in the survey.
    }
\end{figure}

Users were from approximately 100 institutions in 20 countries (specification of country and institution was optional).
The distribution of users across institutions is highly skewed, with only 4 institutions associated with more than 20 answers.
The distribution across countries is skewed as well, with most answers coming from Italy or from an unspecified location, but with the Italian version of the form; we received 193 answers from 19 countries different from Italy and 12 answers from an unspecified country with the English version of the form.
Although the user population is obviously not a statistically significant sample of the full population, we believe the results are indeed relevant and useful.
The breakdown of users depending on their self-described role in the respective institution is given in Figure \ref{fig:user-breakdown}.
It can be seen that the population is not concentrated on any specific role.

\begin{figure}
    \centering
    \includegraphics[width=1\linewidth]{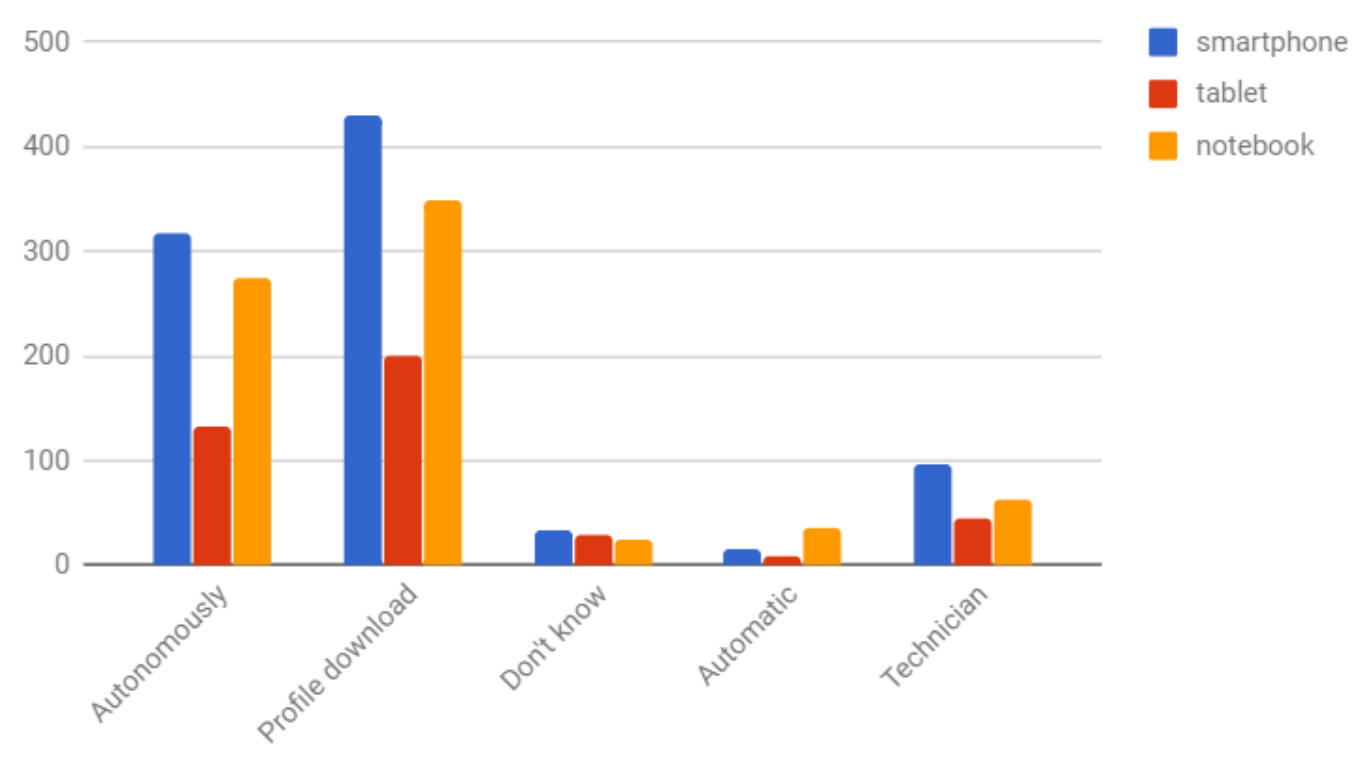}
    \includegraphics[width=1\linewidth]{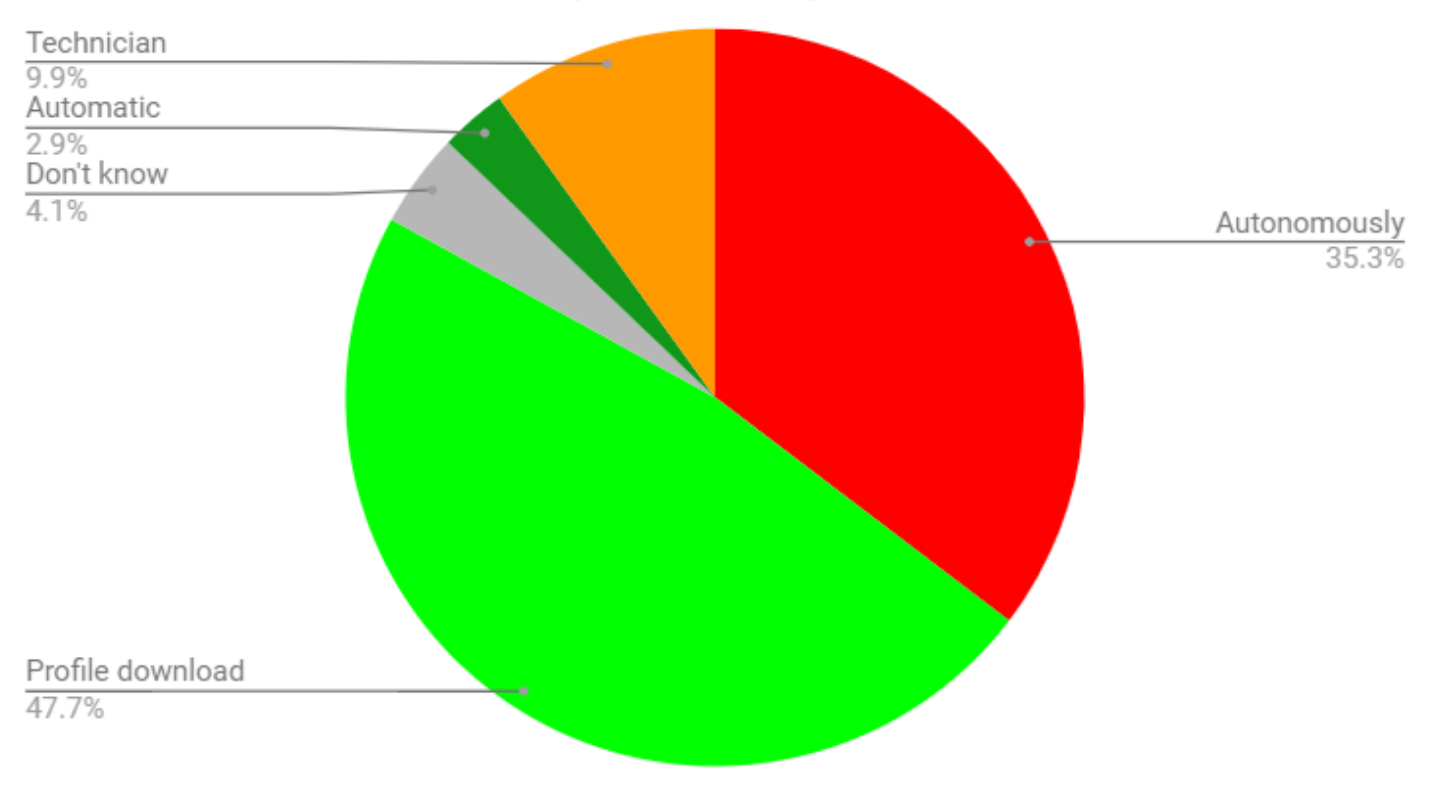}
    \caption{
        \label{fig:configuration-procedure}
        Configuration procedure followed by users. Number of answers for each
        category of devices (up), percentage of answers all categories (down).
    }
\end{figure}

The key question was \emph{``how did you configure your device for connecting to eduroam?''}, which was formulated separately for smartphone, tablet and notebook. The possible answers were:
\begin{enumerate}
    \item \emph{``My device is automatically managed by my organization'';}
    \item \emph{``I have downloaded and installed the network profile of my organization'';}
    \item \emph{``I have played with configuration options until it worked'';}
    \item \emph{``A technician of my organization has configured my device for me'';}
    \item \emph{``I don't know''.}
\end{enumerate}
Mapping from these answers to the security of the resulting configuration is not automatic. We considered the possibility of defining different answers that could have facilitated this mapping but quickly rejected this option: the number of possible answers would have been very large; the survey would have been suitable only for skilled users; and, filling the form would have been quite cumbersome and time consuming. Thus, we preferred more general answers in order to collect as many users as possible. Said this, we believe it is reasonable to assume that:
\begin{itemize}
\item categories 1 and 2 correspond to a secure configuration;
\item categories 3 and 4 may or may not correspond to a secure configuration (these two categories will be analyzed in more detail below)
\end{itemize}
The answers are summarized in Figure \ref{fig:configuration-procedure}.

It can be seen that near half of the supplicants have been configured with the network profile published by the respective organization.
This result is certainly highly positive from the point of view of the developers of the tool for automatic configuration (eduroam-cat) but, in our opinion, is highly worrying: only half of the supplicants are certainly configured securely. We believe that this result is a very strong indication that the underlying hypothesis in WPA2 Enterprise (i.e., that supplicants are configured correctly) is not realistic.

\begin{figure}
    \centering
    \includegraphics[width=1\linewidth]{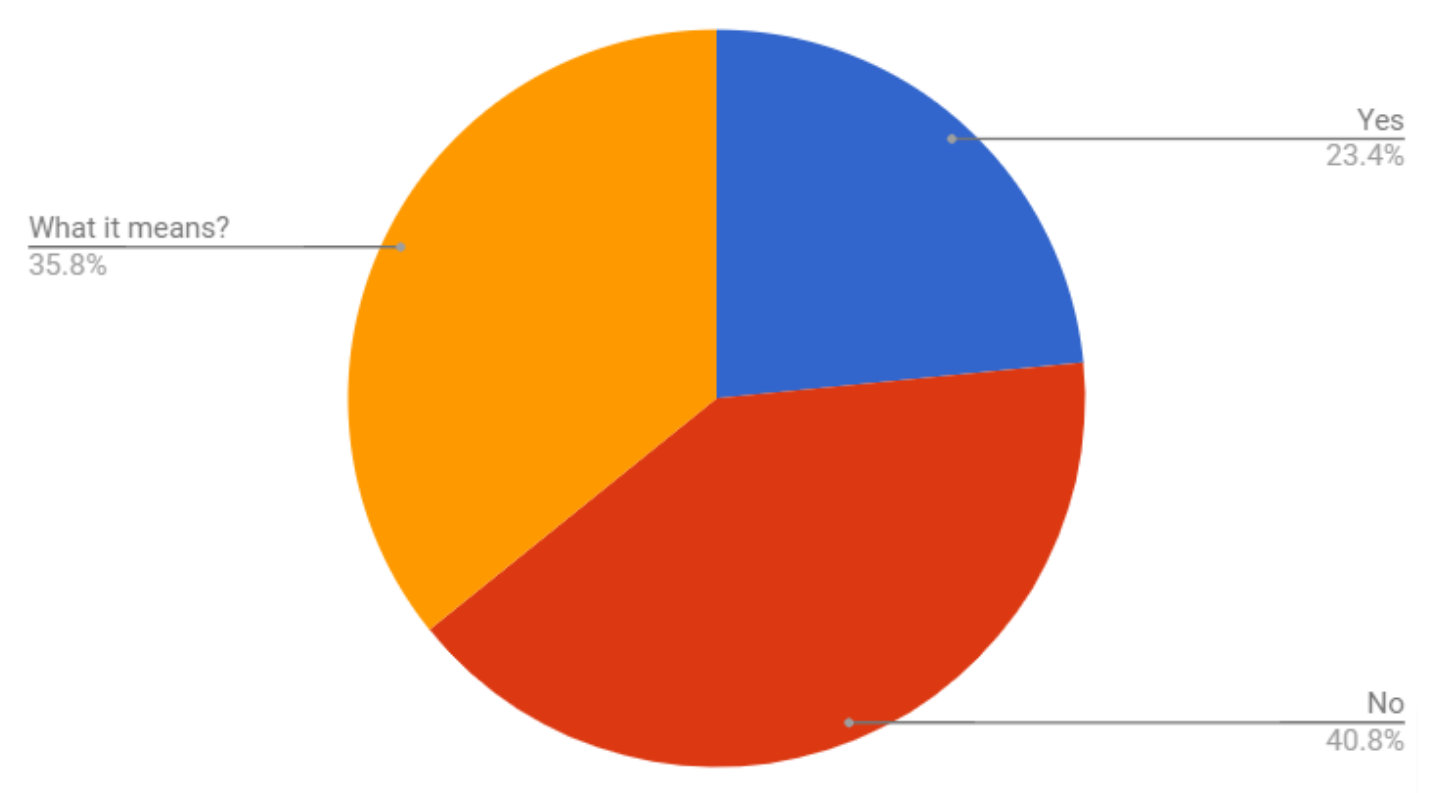}
\caption{
        \label{fig:do-you-know-as}
        Answers of users to the question "do you know the name of the correct authorization server for your network?"
    }
\end{figure}

Concerning category 3 \emph{``I have played with configuration options until it worked''} (35.3\% of devices), we have no elements for conclusively determining whether the resulting configuration was secure or not.
However, there are strong signals suggesting that for many users this is not the case.
Figure \ref{fig:do-you-know-as} summarizes the answers to the question \emph{``do you know the name of the eduroam Authorization Server of your organization?''} given by users that have connected at least one of their devices to eduroam.
It can be seen that 76.6\% of the users either do not know that name or do not even know what an Authorization Server is.
Although knowledge of the name of the Authorization Server is, strictly speaking, independent of the security of the resulting configuration, it seems reasonable to assume that if users in these categories proceed with trial and error until connecting, then the resulting supplicant configuration is unlikely to be secure.

Concerning, finally, category 4 \emph{``A technician of my organization has configured my device for me''} (9.9\% of devices), it would be tempting to assume that all these devices have been configured securely, but such an assumption is probably excessively optimistic.
To our surprise, we have found several eduroam configuration guides published on the web that suggest insecure supplicant configurations, the most frequent insecure suggestion being ``do not validate server certificate''.
Thus, either technicians at those institutions are not fully aware of the security risks, or choose to ignore those risks.
In either case, one cannot take for granted that a technician will configure a supplicant securely.
In order to gain more insights into this important issue, we have performed a more systematic analysis and reviewed a number of configuration guides, as explained in the next section. 
Irrespective of the security of the resulting configuration, though, the fact that less than 13\% of users had the configuration either performed by a technician or managed automatically, implies that supplicant configuration of WiFi Enterprise networks should be designed with unskilled users in mind.

\begin{figure}
    \centering
    \includegraphics[width=1\linewidth]{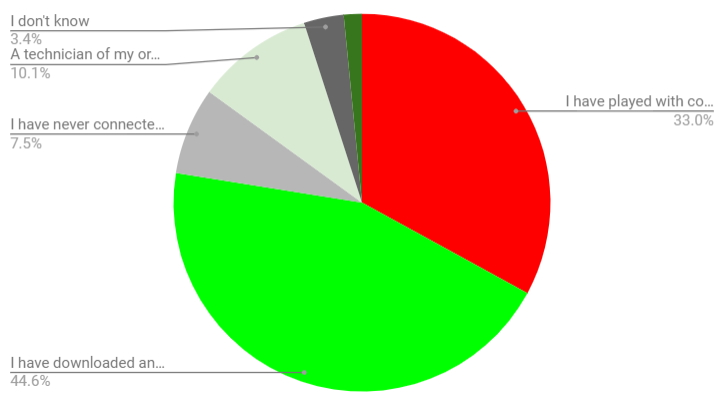}
    \includegraphics[width=1\linewidth]{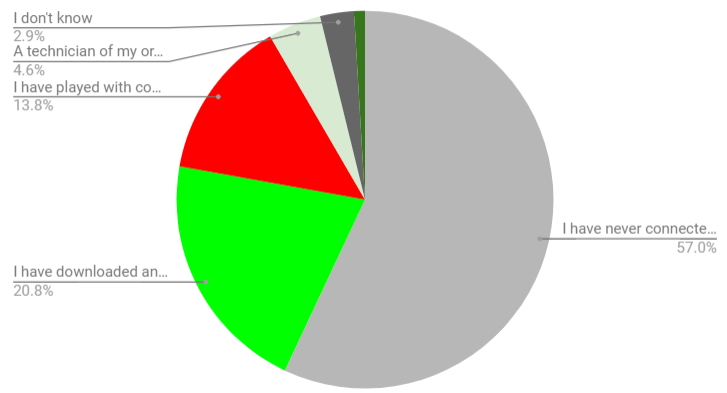}
    \includegraphics[width=1\linewidth]{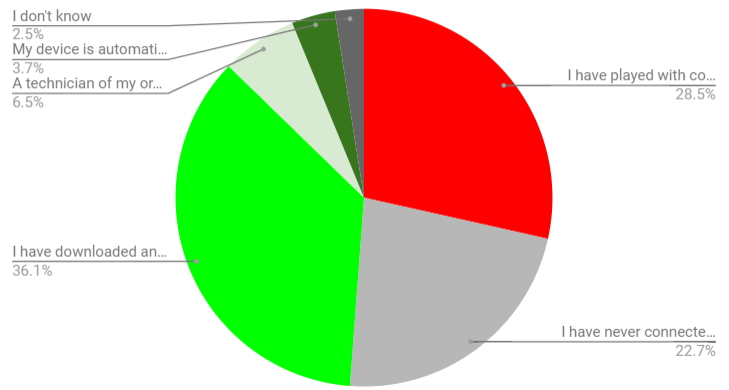}
    \caption{
        \label{fig:configuration-breakdown}
        Configuration procedure followed by users, separately for each device category: smartphone (up), tablet (middle), notebook (down).
    }
\end{figure}

The previous data are relative to the whole set of devices.
The corresponding data for each of the three device categories---smartphone, tablet, notebook---are given in Figure~\ref{fig:configuration-breakdown}.
These data provide an important insight: smartphones are proportionally much more likely to be connected to eduroam than either tablets or notebooks---the percentage of devices that have never been connected to eduroam are 7.5\%, 57\% and 22.7\%, respectively.
This result confirms that smartphones are a very attractive target for attackers, more so than either tablets or notebooks: smartphones are more likely to contain enterprise credentials; and, smartphones may be attacked by an evil twin everywhere outside of the enterprise, without any need of requiring any specific action from the user.

Another observation that can be made from Figure~\ref{fig:configuration-breakdown} is that the distribution of configuration procedures for smartphones is roughly similar to the distribution for all devices (Figure \ref{fig:configuration-procedure}), more so than in the case of either tablets or notebooks.
In particular, we remark than one third of users that connect their smartphone to eduroam do so by playing with configuration options.

Our data confirm earlier analyses executed on a smaller and more localized scale.
An experiment in which the configuration of 507 devices of approximately 350 users in the same University was checked is reported in~\cite{Brenza2015-mm} (along with a more detailed analysis of potential vulnerabilities in eduroam deployments, along with suggestions for possible mitigations).
The users were explicitly invited to participate in an event for checking the configuration of their devices and had to go to a room prepared for this purpose; thus, it is fair to assume that the involved users were more technically informed and security-savvy than the average of the entire user population.
Furthermore, the event was organized five months after updating the eduroam configuration guide and after distributing specific leaflets.
Despite these very favorable conditions, it turned out that 52\% of the devices were wrongly configured.
A report summarizing a 4-year experience of deploying WPA2 Enterprise in an University (not participating in eduroam) reported that 43\% of their students ``do not use CA certificates to authenticate the server''~\cite{Yanson2016-fh}.

We cannot compare our results to data obtained in corporate environments, as we could not find any such data.
We are only aware of surveys in large US companies focussing on policies for enterprise usage of personal devices (\emph{bring your own device, BYOD}): interestingly, half of the organizations that support or mandate usage of personal devices do not have a formal policy for their usage \cite{noauthor_2016-vz,noauthor_2016-vc}.
Furthermore, half of the respondents to another similar survey use their personal devices in the workplace even in organizations that explicitly ban the use of those devices, or in the absence of any specific policy or without even knowing whether such a policy exists \cite{Jones2012-jt}.
These facts are a further proof of the relevance of the issues that we are considering here, we believe.

\section{Survey of configuration guides}
We reviewed 311 configuration guides published on the web by 69 institutions in 17 countries (each guide being for a specific supplicant operating system).
We are not aware of any similar analysis.
The institutions are a subset of the institutions found in the user survey of the previous section (we could not afford to analyze all the institutions).
We considered only configuration guides either in English or in Italy.

We placed each guide into a single category, depending on whether the guide contained any indication that could lead to an insecure supplicant configuration. The categories are as follows:
\begin{description}
\item[Do not validate server certificate] The guide specifies to uncheck the ``validate server certificate'' option.
\item[No validation info] The guide does not provide any information about the need to validate the certificate, nor about the certificate itself.
\item[Any certificate] The guide specifies explicitly to accept any certificate.
\item[No certificate info] The guide specifies that a certificate has to be accepted, but it does not provide any indication about the certificate itself (i.e., about which certificate should be accepted).
\item[Incomplete certificate info] The guide shows the server name (usually in images used for clarifying the idescription) but it does not specify that only a certificate with that exact name should be accepted.
\item[None]The guide does not indicate any action that could lead to an insecure configuration.
\end{description}

\begin{figure}
    \centering
    \includegraphics[width=1\linewidth]{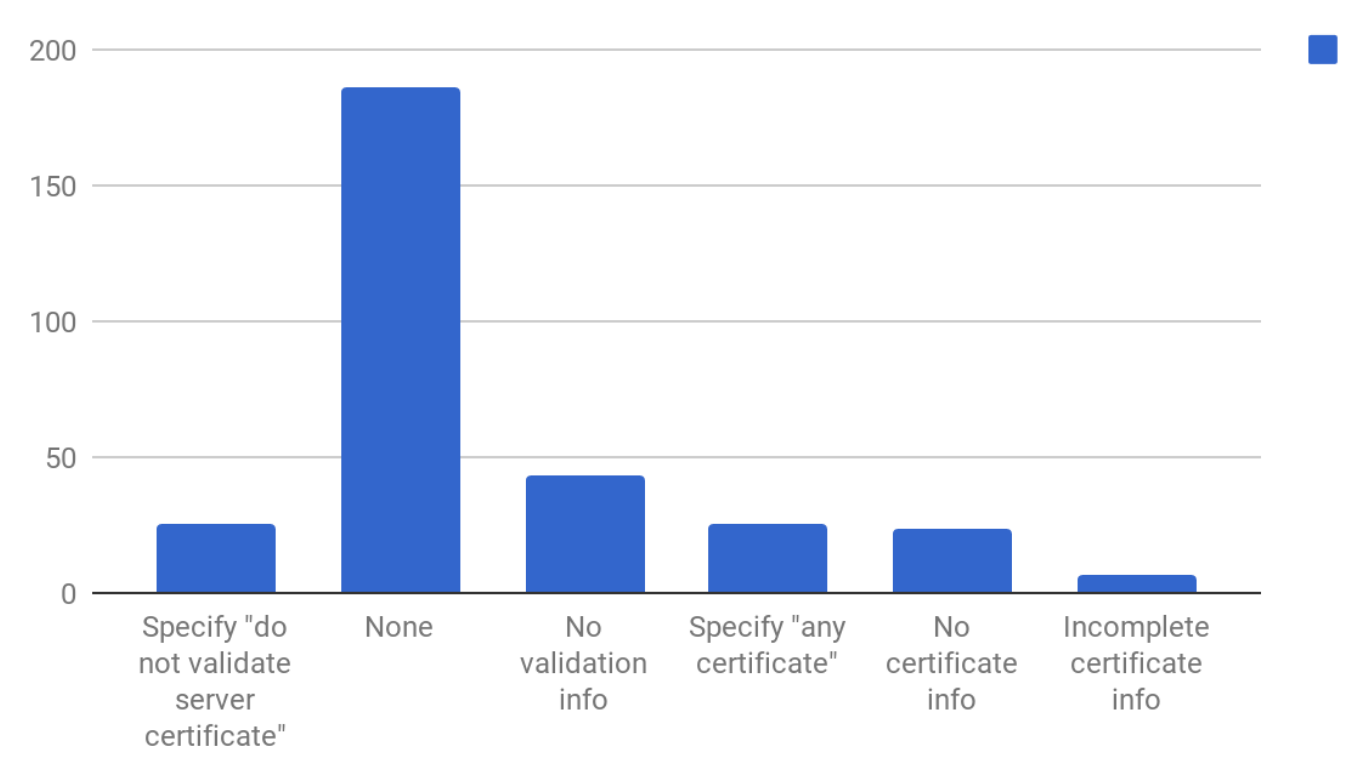}
    \includegraphics[width=1\linewidth]{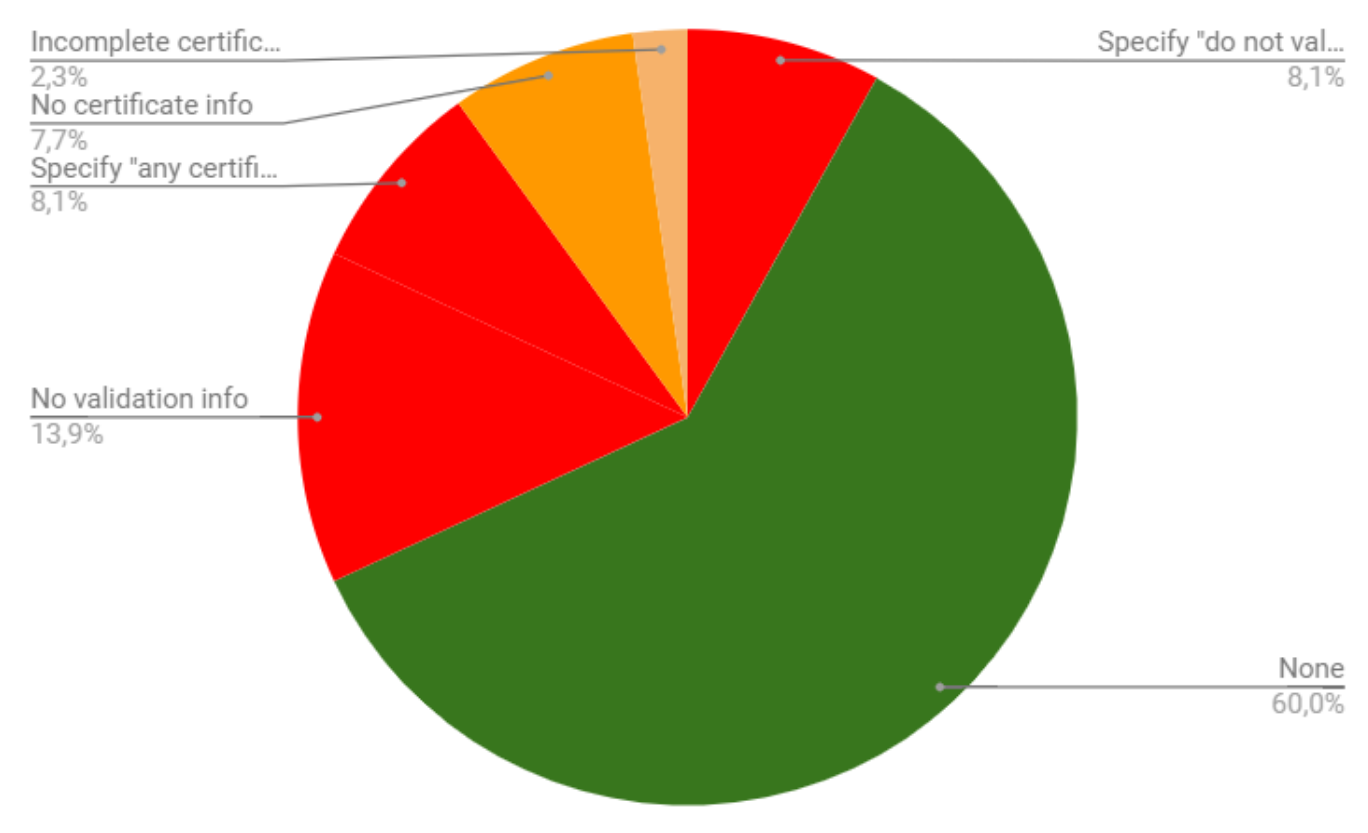}
    \caption{
        \label{fig:categorization-hist}
        Configuration guides by category.
    }
\end{figure}

The distribution of guides by category is shown in Figure~\ref{fig:categorization-hist}.
It can be seen that only 60\% of the guides belong to category None, thus 40\% of the guides contain indications that could lead to an insecure configuration.

The risk levels of the chosen categories are different, thus one might claim that not all the resulting configurations can be considered as being insecure.
In particular, the risk level for guides in categories ``No certificate info'' or ``Incomplete certificate info'' is arguably smaller than for guides in the other categories.
However, even if we consider guides in those categories as suggesting a \emph{partly secure} configuration, it is simple to realize from Figure~\ref{fig:categorization-hist} that these guides would account for just 10\% of all guides: the percentage of guides suggesting an insecure supplicant configuration would still be around 30\%.

To ascertain whether this result is an artifact of a few institutions publishing many guides with questionable indications, we counted how many institutions published only guides describing a secure configuration.
We found that only 29 institutions published guides of this kind, while the remaining 40 institutions published at least one guide leading to a configuration that is either insecure or partly secure (in the meaning above). The vast majority of these institutions (36 out of 40) published at least one insecure guide.

This result is a strong indication that, as claimed in the previous sections, one cannot take for granted that a technician will configure a supplicant securely. We believe this result is highly significant and constitute a further proof that a fundamental security assumption in WPA2 Enterprise---supplicants must be configured appropriately---is no longer realistic. 

Of course, by no means we are implying that technicians that prepared the guides in our analysis are not competent.
We are merely describing the existing scenario.
There may be many reasons why a configuration guide indicates steps that could lead to an insecure configuration, for example, the need of prioritizing the effort of the (usually scarce) IT staff could have led to the decision of writing configuration guides that are simpler for users to read and for technicians to maintain, even at the expense of an increased risk, in exchange for more time to devote to other activities for increasing the overall security of the organization.
Indeed, we are aware of several institutions of the eduroam federation that carefully configured their infrastructures so that supplicants may connect only by using the network profile installed by the eduroam-cat application \footnote{This property is obtained by a smart exploitation of the fact that WPA2 Enterprise protocols allow using two different identities during an execution of the authentication protocol; these infrastructures are configured to require an \emph{outer-identity} identical for all users, hard-coded within the network profile of eduroam-cat and unlikely to be guessed by trial and error; and an \emph{inner-identity} specific for each user.}.

Finally, we observed that 47 of the 69 institutions make a network profile available for downloading and usage with eduroam-cat.
We could not identify any significant signal linking the possibility of using eduroam-cat and the quality of the configuration guides.
Interestingly, we observed that those configuration guides that mentioned eduroam-cat, tended to describe this tool as a mean for \emph{simplifying} the configuration; the fact that eduroam-cat allows obtaining a \emph{secure} configuration is mentioned much less frequently, if at all.

\section{Conclusions}
Techniques for attacking WPA2 Enterprise supplicants that are not configured appropriately are widely known (e.g., \cite{Yavor_undated-kk,noauthor_2013-gh,Cassola2013-ox,Snoodgrass_J_undated-kf,Brenza2015-mm}).
These techniques are not a weakness of the WPA2 Enterprise family of protocols: they are an obvious consequence of using those protocols without satisfying their basic requirements.
The corresponding risks have become much more pervasive and significant than they used to be, though, and it seems fair to claim than neither users nor technicians have realized the magnitude of these risks.
The world is now very different from how it was when WPA2 Enterprise was defined, in 2004: today virtually every user carries a WiFi-enabled personal device that attempts to connect to known networks permanently and automatically, without any explicit action from the user; what is worse, such devices often connect by using credentials that unlock access to all services of an enterprise.
The combination of these two factors makes attacks aimed at stealing enterprise credentials from smartphones extremely attractive \cite{Bartoli2018-pd}.

In this work we conducted two extensive analyses aimed at assessing how pervasive the corresponding risks are.
A survey among approximately 1000 users from many different institutions and a review of approximately 310 network configuration guides. We are not aware of any similar study.
Our analyses provide important insights into the practical deployments of such a pervasive and fundamental technology as WPA2 and strongly suggest that a requirement fundamental for enjoying the security guarantees of WPA2 Enterprise is not satisfied in a very large amount of deployments.

We hope that our results will help in disseminating awareness of the risks associated with insecure configuration of WPA2 Enterprise supplicants and in promoting more secure usage and configuration practices.
We also hope that our results will be taken into account by the WiFi Alliance for defining the new family of wireless protocols WPA3.

\subsection*{Acknowledgments}

We are grateful to all the people which helped disseminate the survey and to all those which answered.
We are grateful to Eleonora Girardini for her help in collecting and analyzing the configuration guides and to Stefan Winter for his comments on the results of the survey.



\bibliographystyle{plain}
\bibliography{ms} 

\end{document}